\documentclass[floats,prb,aps,twocolumn]{revtex4}
\usepackage{graphicx}
\begin{document}


\title{Bose-Einstein condensation and superfluidity of magnetobiexcitons in quantum wells' and graphene superlattices}

\author{Oleg L. Berman$^{1}$,  Roman Ya. Kezerashvili$^{1}$   and       Yurii E. Lozovik$^{2}$}

\affiliation{\mbox{$^{1}$Physics Department, New York City College of Technology, the City University of New York,}
\\ Brooklyn, NY 11201, USA
 \\ \mbox{$^{2}$ Institute of Spectroscopy, Russian Academy of
Sciences,}  \\ 142190 Troitsk, Moscow Region, Russia}


\begin{abstract}
We propose the Bose-Einstein condensation (BEC) and superfluidity of  quasi-two-dimensional (2D)
spatially indirect magnetobiexcitons in  a slab of superlattice with
alternating electron and hole layers  consisting from the semiconducting quantum wells (QWs) and graphene superlattice
in high magnetic field. The two different Hamiltonians of
a dilute gas of magnetoexcitons with a dipole-dipole repulsion in superlattices consisting of both QWs and graphene layers in the limit of high magnetic field
have been reduced  to one effective Hamiltonian a dilute gas of
two-dimensional excitons without magnetic field.  Moreover, for $N$ excitons we have reduced the problem of $2N\times 2$ dimensional space onto the problem of $N\times 2$  dimensional space by integrating over the coordinates of the relative motion of an electron (e) and a hole (h). The instability of the ground state of the system of interacting
two-dimensional indirect magnetoexcitons in a slab of superlattice with
alternating electron and hole layers  in high magnetic field is established. The stable
system of  indirect quasi-two-dimensional magnetobiexcitons, consisting
of pair of indirect excitons with opposite dipole moments is
considered. The density of superfluid
component $n_{s}(T)$ and
 the temperature of the Kosterlitz-Thouless phase transition to the superfluid state in
the  system of  two-dimensional indirect magnetobiexcitons, interacting as
electrical quadrupoles, are obtained for both QW and graphene realizations.

\vspace{0.3cm}


\end{abstract}

\maketitle


The many-particle systems of the spatially-indirect excitons in coupled
quantum wells (CQWs) in the presence or absence of a magnetic field $B$ have
been the subject of recent experimental studies. \cite{Snoke,Butov,Eisenstein} These systems are of interest, in
particular, in connection with the possibility of BEC and superfluidity of indirect excitons or electron-hole pairs, which
would manifest itself in the CQWs as persistent electrical currents in each
well and also through coherent optical properties and Josephson phenomena
\cite{Lozovik,Birman,Littlewood,Vignale,Berman}. In strong magnetic fields ($%
B>7$T) 2D excitons survive in a substantially wider
temperature region, as the exciton binding energies increase with magnetic
field\cite{Lerner,Paquet,Kallin,Yoshioka,Ruvinsky,Ulloa,Moskalenko}. The
problem of essential interest is also collective properties of
magnetoexcitons in high magnetic fields in superlattices and layered system \cite{Filin}.

Recent technological advances have allowed the production of graphene, which
is a 2D honeycomb lattice of carbon atoms that form the basic planar
structure in graphite \cite{Novoselov2,Zhang2}. Graphene has been attracting
a great deal of experimental and theoretical attention because of unusual
properties in its bandstructure \cite{Novoselov2,Zhang2,Nomura,Jain}. It is
a gapless semiconductor with massless electrons and holes which have been
described as Dirac-fermions \cite{DasSarma}. Since there is no gap between
the conduction and valence bands in graphene without magnetic field, the
screening effects result in the absence of excitons in graphene in the
absence of a magnetic field. A strong magnetic field produces a gap since
Landau levels form the discrete energy spectrum. The gap reduces screening
and leads to the formation of magnetoexcitons. We consider magnetoexcitons
in the superlattices with alternating electronic ($e$) and holes ($h$) QWs as
well as graphene layers (GLs) and suppose that recombination times may be much
greater than relaxation times $\tau _{r}$ due to small overlapping of the
spatially separation of e- and h- wave functions in QWs or GLs.
In this case electrons and holes are characterized by different
quasiequilibrium chemical potentials. Therefore, in the system of indirect
excitons in superlattices the quasiequilibrium phases appear,
as, for example, in CQWs \cite{Lozovik,Berman}. While coupled-well structures
with spatially separated electrons and holes are typically considered to be
under applied electric field, which separates electrons and holes in
different QWs \ \cite{Butov}, we assume there are no
external fields applied to a slab of superlattice. If electron and hole
QWs alternate, there are excitons with parallel dipole moments in
one pair of wells, but dipole moments of excitons in other neighboring
pairs of neighboring wells have opposite direction. This fact leads to
essential distinction of properties of $e-h$ system in superlattices from
one for CQWs with spatially separated electrons and holes,
where indirect exciton system is stable due to a dipole-dipole repulsion of
all excitons. This difference manifests itself already beginning from
three-layer $e-h-e$ or $h-e-h$ system.  The superlattice of the GLs with the electrons and holes separated in alternating layers
can be created by controlling  the chemical potential of the charge carriers due to the doping of the GLs by the charged impurities \cite{Schedin}.

In this Letter we propose a new physical realization of magnetoexcitonic BEC
and superfluidity in superlattices with alternating electronic and hole
layers, that is in a sense representing an array of QWs or GLs
with spatially separated electrons and holes in high magnetic field. We
reduce the problem of magnetoexcitons to the problem of excitons at $B=0$.
The instability of the ground state of the system of interacting indirect
excitons in slab of superlattice with alternating $e-$ and $h-$ layers is
established in a strong magnetic field. Two-dimensional indirect
magnetobiexcitons, consisting of the indirect magnetoexcitons with opposite
dipole moments, are considered in a high magnetic field. These
magnetobiexcitons repel as electrical quadrupoles at long distances. As a
result, the system of indirect magnetobiexcitons becomes stable. Below the radius
and the binding energy of indirect magnetobiexciton are calculated. By
apllying the ladder approximation we consider a collective spectrum of the
weakly interacting by the quadrupole law two-dimensional indirect
magnetobiexcitons and calculate their superfluid density $n_{s}(T)$ in
superlattices at low temperatures $T$. We analyze the dependence of the
Kosterlitz-Thouless transition\cite{Kosterlitz} temperature and superfluid
density on magnetic field.

The Hamiltonian of a single 2D magnetoexciton  is given in Refs.~[\onlinecite{Lerner,Kallin}] for CQWs  and in Ref.~[\onlinecite{Iyengar}] for GLs.
A conserved quantity for an isolated electron-hole pair in magnetic field $B$
is the exciton generalized momentum $\hat{\mathbf{P}}$ defined as $\hat{\mathbf{P}}=-i\hbar \nabla _{e}-i\hbar \nabla _{h}+e/c(\mathbf{A%
}_{e}-\mathbf{A}_{h})-e/c[\mathbf{B}\times (\mathbf{r}_{e}-\mathbf{r}%
_{h})]$, for the Dirac equation in GLs\cite{Iyengar} as well as for the
Schr\"{o}dinger equation in CQWs\cite{Gorkov,Lerner,Kallin}. Here
$\mathbf{r}_{e}$ and $\mathbf{r}_{h}$ are electron and hole
locations along QWs, respectively, $D$ is the distance between
electron and hole QWs, $e$ is the charge of an electron, $c$ is
the speed of light and $\epsilon $ is a dielectric constant.  We find it is convenient to work with
the symmetric gauge for a vector potential $\mathbf{A}_{e(h)}=1/2[\mathbf{B}%
\times \mathbf{r}_{e(h)}]$, where $\mathbf{r}_{e(h)}$ is the location of an electron or a hole, respectively.

The Hamiltonian of a single isolated magnetoexciton without any random field
($V_{e}(\mathbf{r}_{e})=V_{h}(\mathbf{r}_{h})=0$) is commutated with $\hat{%
\mathbf{P}}$, and hence they have the same the eigenfunctions, which have
the following form (see Refs.~[\onlinecite{Lerner,Gorkov}]):
\begin{equation}
\Psi _{k\mathbf{P}}(\mathbf{R},\mathbf{r})=e^{\left\{ i\mathbf{R}\left(
\mathbf{P}+\frac{e}{2}\mathbf{B}\times \mathbf{R}\right) +i\gamma \frac{%
\mathbf{P}\mathbf{r}}{2}\right\} }\Phi _{k}(\mathbf{P},\mathbf{r}),
\end{equation}%
%
where $\gamma =(m_{h}-m_{e})/(m_{h}+m_{e})$ ($m_{e}$ and $m_{h}$ are the masses of an electron and hole in QWs, and $\gamma = 0$ for GLs), $\Phi _{k}(\mathbf{P},%
\mathbf{r})$ is a function of internal coordinates $\mathbf{r}$ and the
eigenvalue of the generalized momentum  $\mathbf{P}$, and $k$ represents the
quantum numbers of exciton internal motion. The expression for $\Phi _{k}(%
\mathbf{P},\mathbf{r})$ for CQWs is given in Ref.~[%
\onlinecite{Lerner,Ruvinsky}], and in Ref.~[\onlinecite{Iyengar}]) for
graphene layers. In high magnetic fields the magnetoexcitonic quantum
numbers $k=\{n_{+},n_{-}\}$ for an electron in Landau level $n_{+}$ and a
hole in level $n_{-}$.

For large electron-hole separation $D\gg r_{B}$, where $r_{B} = \sqrt{\hbar c/(eB)}$,  transitions between Landau
levels due to the Coulomb electron-hole attraction can be neglected, if the
following condition is valid for the magnetoexcitonic binding energy  $E_{b}$:
 $E_{b}=e^{2}/(\epsilon _{b}D)\ll \hbar \omega _{c}=\hbar
eB(m_{e}+m_{h})/(2m_{e}m_{h}c)$ for QWs; $E_{b}=4e^{2}/(\epsilon D)\ll \hbar
v_{F}/r_{B}$ for GLs, where   $v_{F}=\sqrt{3}at/(2\hbar )$ is the
Fermi velocity of electrons for a lattice constant $a=2.566\mathring{A}$ and
the overlap integral between nearest carbon atoms  $t\approx 2.71eV$ \cite%
{Lukose}. This corresponds to high magnetic field $B$, large interlayer
separation $D$ and large dielectric constant of the insulator layer between
the GLs.  At small densities $nr_{B}^{2}\ll 1$ the system of indirect
excitons at low temperatures is the two-dimensional dilute weakly nonideal Bose gas
with normal to wells dipole moments $\mathbf{d}$ in the ground state ($d\sim
eD$, $D$ is the interwell separation), increasing with the distance between
wells $D$. In contrast to ordinary excitons, for a low-density spatially
indirect magnetoexciton system the main contribution to the energy is
originated from dipole-dipole interactions $U_{-}$ and $U_{+}$ of
magnetoexcitons with opposite and parallel dipoles, respectively. Two
parallel ($+$) and opposite ($-$) dipoles in low-density system interact as $%
U_{+}=-U_{-}=e^{2}D^{2}/\epsilon R^{3}$, where $R$ is the distance between
dipoles along wells planes. We suppose that $D/R\ll 1$ and $L/R\ll 1$ ($L$
is the mean distance between dipoles normal to the wells). We consider the
case, when the number of QWs or GLs $k$ in superlattice is restricted $k \ll \left(D\sqrt{\pi n}\right)^{-1}$, where $n$ is exciton surface density, and this is valid for small $k$ or for sufficiently low exciton density.

Due to the orthonormality of the single magnetoexciton wave functions
$\Phi _{n_{+},n_{-}}(\mathbf{0,r})$ ($\mathbf{r} = \mathbf{r}_{e}-\mathbf{r}_{h}$) the projection of the the many-particle Hamiltonian onto the lowest
Landau level results in the effective Hamiltonian, which does not reflect
the spinor nature of the four-component magnetoexcitonic wave functions in
graphene. Typically, the value of $r$ is $r_{B}$, and $P\ll \hbar
/r_{B}$. In this approximation, the effective Hamiltonian $\hat{H}_{\mathrm{%
eff}}$ in the momentum representation $P$ in the subspace the
lowest Landau level (for QWs $n_{+}=n_{-}=0$; for GLs $%
n_{+}=n_{-}=1$) has the same form (compare with Ref.[\onlinecite{Berman}])
as for two-dimensional boson system without a magnetic field. Only differences are that, instead of the exciton mass $M = m_{e} + m_{h}$ and ordinary momenta, we have the magnetoexciton magnetic mass, which depends on $B$ and $D$ and magnetic momenta, respectively. For the lowest
Landau level we denote the spectrum of the single exciton $\varepsilon
_{0}(P)\equiv \varepsilon _{00}(\mathbf{P})$. The Hamiltonian can be represented in the form: $\hat{H}_{tot}=\hat{H}_{0}+\hat{H}_{int}$, where $\hat{H}_{0}$ is the effective Hamiltonian of the system of noninteracting magnetoexcitons:
\begin{eqnarray}
\label{H0}
\hat{H}_{0}=\sum_{\mathbf{p}}^{{}}\varepsilon _{0}(p)(a_{\mathbf{p}}^{+}a_{%
\mathbf{p}}^{{}}+b_{\mathbf{p}}^{+}b_{\mathbf{p}}^{{}}+a_{-\mathbf{p}%
}^{+}a_{-\mathbf{p}}^{{}}+b_{-\mathbf{p}}^{+}b_{-\mathbf{p}}^{{}}).
\end{eqnarray}
In Eq.~(\ref{H0}) $\varepsilon _{0}(p)=p^{2}/(2m_{B})$ is the spectrum of isolated
two-dimensional indirect magnetoexciton; $\mathbf{p}$ represents the
excitonic magnetic momentum and $a_{\mathbf{p}}^{+}$, $a_{\mathbf{p}}^{{}}$,
$b_{\mathbf{p}}^{+}$, $b_{\mathbf{p}}^{{}}$ are creation and annihilation
operators of magnetoexcitons with up and down dipoles. For an isolated magnetoexciton on the lowest Landau level at the small
magnetic momenta under consideration, $\varepsilon _{0}(\mathbf{P})\approx
P^{2}/(2m_{B})$, where $m_{B}$ is the effective \textit{magnetic} mass of a
magnetoexciton in the lowest Landau level and is a function of the distance $%
D$ between $e$ -- and $h$ -- layers and magnetic field $B$ \ \cite{Ruvinsky}. In a strong magnetic field at $D\gg r_{B}$ the
exciton magnetic mass is $m_{B}(D)=\epsilon D^{3}/(e^{2}r_{B}^{4})$ for QWs
\cite{Ruvinsky} and $m_{B}(D)=\epsilon D^{3}/(4e^{2}r_{B}^{4})$ for GLs \cite{Berman_Lozovik_Gumbs}. The effective
Hamiltonian of the interaction between magnetoexcitons $\hat{H}_{int}$ is:
\begin{equation}
\begin{array}{c}
\label{hint}\hat{H}_{int}=\frac{U}{2S}\sum_{\mathbf{p}_{1}+\mathbf{p}_{2}=%
\mathbf{p}_{3}+\mathbf{p}_{4}}^{{}}(a_{\mathbf{p}_{4}}^{+}a_{\mathbf{p}%
_{3}}^{+}a_{\mathbf{p}_{2}}a_{\mathbf{p}_{1}}+\nonumber \\
b_{\mathbf{p}_{4}}^{+}b_{\mathbf{p}_{3}}^{+}b_{\mathbf{p}_{2}}b_{\mathbf{p}%
_{1}}-a_{\mathbf{p}_{4}}^{+}a_{\mathbf{p}_{3}}^{+}b_{\mathbf{p}_{2}}b_{%
\mathbf{p}_{1}}-b_{\mathbf{p}_{4}}^{+}b_{\mathbf{p}_{3}}^{+}a_{\mathbf{p}%
_{2}}a_{\mathbf{p}_{1}}-\nonumber \\
a_{\mathbf{p}_{4}}^{+}b_{\mathbf{p}_{3}}^{+}a_{\mathbf{p}_{2}}b_{\mathbf{p}%
_{1}}),%
\end{array}%
\end{equation}%
where $S$ is the surface of the system. Let us consider the temperature $T=0$.
We apply Bogolubov approximation and assume $(N-N_{0})/N_{0}\ll 1$,
where $N$ and $N_{0}$ are the total number of particles and the number of
particles in condensate, respectively. In Bogolubov approximation the interaction between non-condensate
particles is neglected, and only the interactions between condensate particles and excited particles with
condensate particles are considered. Therefore, in the total Hamiltonian the terms, arising from first and second terms of the
Hamiltonian (\ref{hint}), which describe the repulsion of the indirect
magnetoexcitons with parallel dipole moments, are compensating by other
terms of this Hamiltonian, describing the attraction of
indirect magnetoexcitons with opposite dipoles. As the result only terms
describing the attraction survive. Let us diagonalize  the total Hamiltonian  by using
of the Bogolubov type unitary transformation
\begin{equation}
\begin{array}{c}
\label{bog}a_{\mathbf{p}}=\frac{1}{\sqrt{1-A_{\mathbf{p}}^{2}-B_{\mathbf{p}%
}^{2}}}(\alpha _{\mathbf{p}}+A_{\mathbf{p}}\alpha _{-\mathbf{p}}^{+}+B_{%
\mathbf{p}}\beta _{-\mathbf{p}}^{+}),\nonumber \\
b_{\mathbf{p}}=\frac{1}{\sqrt{1-A_{\mathbf{p}}^{2}-B_{\mathbf{p}}^{2}}}%
(\beta _{\mathbf{p}}+A_{\mathbf{p}}\beta _{-\mathbf{p}}^{+}+B_{\mathbf{p}%
}\alpha _{-\mathbf{p}}^{+}),%
\end{array}%
\end{equation}%
where the coefficients $A_{\mathbf{p}}$ and $B_{\mathbf{p}}$ are found from
the condition of vanishing of coefficients at nondiagonal terms in the
Hamiltonian.
 As the result we obtain $\hat H_{tot} = \sum_{{\bf p} \ne 0}^{}  \varepsilon (p)
(\alpha _{{\bf p}}^{+}\alpha _{{\bf p}}^{} +
\beta _{{\bf p}}^{+}\beta _{{\bf p}}^{})$
with  the  spectrum of
quasiparticles $\varepsilon (p)$:
$\varepsilon (p) = \left[(\varepsilon_{0}(p))^{2} - (nU)^{2}\right]^{1/2}$.
At small momenta $p<\sqrt{2m_{B}nU}$ the spectrum of excitations becomes
imaginary. Hence, the system of weakly interacting indirect magnetoexcitons
in a slab of superlattice is instable. The instability of magnetoexcitons
becomes stronger as magnetic field higher, because $m_{B}$ increases with
the increment of magnetic field, and, therefore, the region of $p$
resulting in the imaginary collective spectrum increases as $B$ increases.

This, on first view, strange result can be illustrated by the following
example. There are equal number of dipoles oriented up and down. Let us
consider four dipoles, two of them being oriented up and two --- down. It is
easy to count that number of repelling pairs is smaller than that of
attracting ones. The prevailing of attraction leads to instability.

 We assume the energy degeneracy respect to two
possible spin projections in QWs and graphene and two graphene
valleys (two pseudospins). Since electrons on a graphene lattice can be in
two valleys, there are four types of excitons in bilayer graphene. Due to
the fact that all these types of excitons have identical envelope wave
functions and energies\cite{Iyengar}, we consider below only excitons in one
valley. Also, we use $n_{0} = n/(4s)$ as the density of excitons in graphene
superlattice, with $n$ denoting the total density of excitons, and $s$ the
spin degeneracy, which equals to $4$ for magnetoexcitons in bilayer graphene.
Besides, we use $n_{0} = n/s$ as the density of excitons in QWs,
 with the spin degeneracy $s$, which equals to $4$ for magnetoexcitons in QWs.

\begin{figure}
\includegraphics[width=1.2in,height = 3cm]{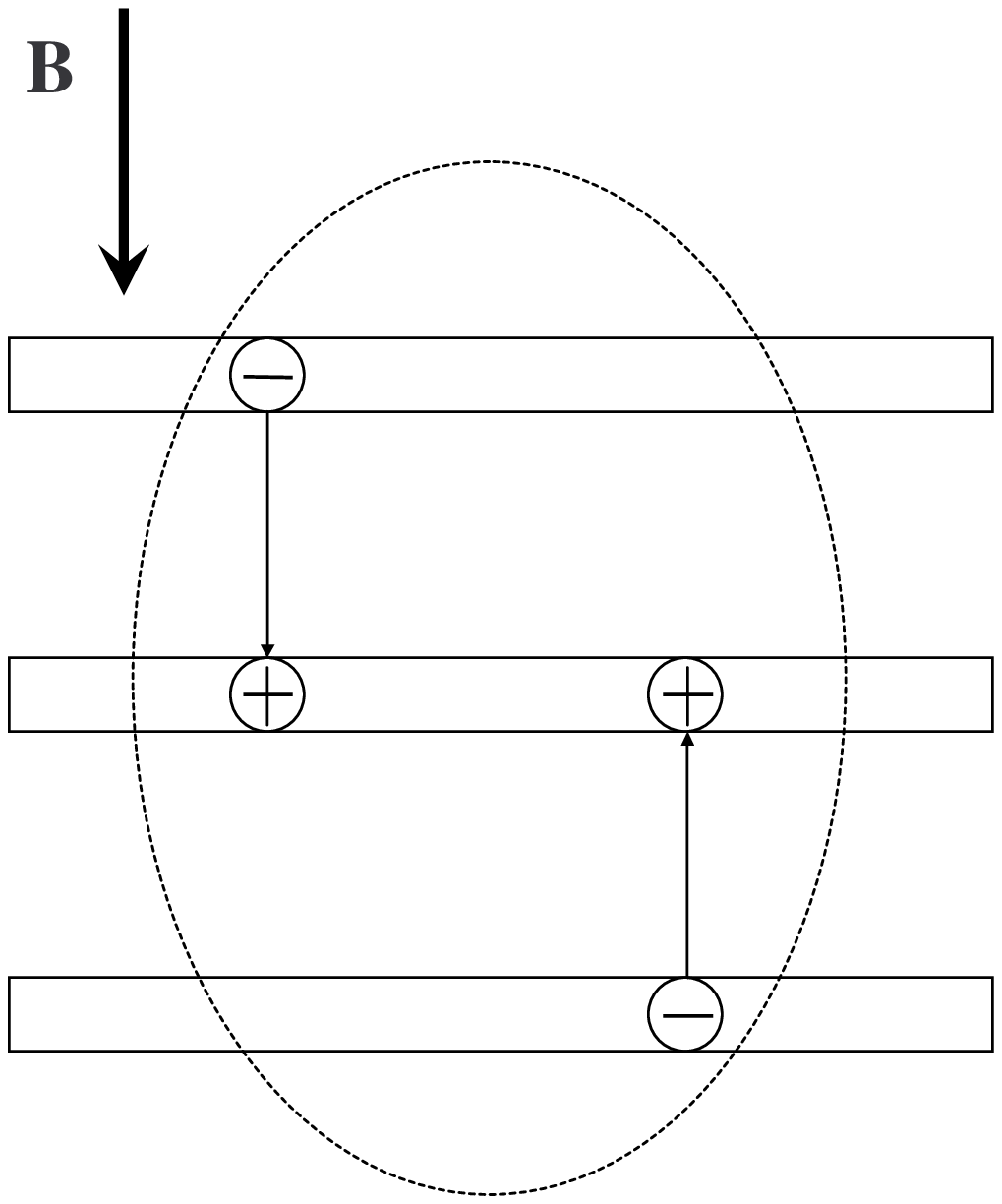}
\caption{2D indirect magnetobiexcitons consisting of indirect
magnetoexcitons with opposite dipole moments, located in neighboring pairs
of QWs (GLs).}
\label{biexciton}
\end{figure}

Let us consider as the ground state of the system the low-density weakly
nonideal gas of two-dimensional indirect magnetobiexcitons, created by
indirect magnetoexcitons with opposite dipoles in neighboring pairs of wells
(Fig.~\ref{biexciton}). The small parameter for the adiabatic approximation
is the numerical small parameter which is equal to the ratio of magnetobiexciton
and magnetoexciton energies or the ratio between radii of magnetoexciton and
magnetobiexciton along QWs or GLs. These parameters are small, and they are even smaller
than analogous parameters for atoms and molecules. The smallness of these
parameters will be verified below by the results of the calculation of
indirect magnetobiexciton. Here it was assumed, that the distance between
wells (GLs) $D$ is greater than the radius of indirect
magnetobiexciton $a_{b}$.  The potential energy of
interaction between indirect magnetoexcitons with opposite dipoles $U(r)$
has the form ($r$ is the distance between indirect
magnetoexcitons along QWs/or GLs):
\begin{eqnarray}  \label{potbi}
U(r) = \frac{e^{2}}{\epsilon r} - \frac{2e^{2}}{\epsilon \sqrt{r^{2} + D^{2}}%
} + \frac{e^{2}}{\epsilon \sqrt{r^{2} + 4 D^{2}}} .
\end{eqnarray}

 At $r > 1.11 D$
indirect magnetoexcitons attract, and at $r < 1.11 D$  they repel.
The minimum of potential energy
$U(r)$ locates at
 $r = r_{0} \approx 1.67 D$ between indirect excitons.
So at large $D$ magnetobiexciton levels correspond to the two-dimensional harmonic
oscillator with the frequency
$\omega = 0.88 e^{2}/(m_{B} \epsilon D^{3})$:
$E_{n} = - 0.04 e^{2}/(\epsilon D) +
 2\sqrt{2}E_{0}\left(r^{*}/D\right)^{3/2}(n + 1)$
where $E_{0} = m_{B}e'^{4}/(\hbar ^{2}\epsilon) $,
$r^{*} = \hbar ^{2}\epsilon/(2m_{B}e'^{2})$, $e'^{2} = 0.88 e^{2}$.
In the ground state the characteristic spread
 of magnetobiexciton $a_{b}$ along QWs/GLs
 (near the mean radius of magnetobiexciton
$r_{0}$ along wells/GLs) is:
$a_{b} = \left[2\hbar/(m_{B}\omega)\right]^{1/2} = (8r^{*})^{1/4}D^{3/4}
= 1.03 a_{ex}$  ,
where $a_{ex} = (8r^{ex})^{1/4}D^{3/4}$;
$r^{ex} = \hbar ^{2}\epsilon/(2m_{B}e^{2})$ is the two-dimensional
effective Bohr radius with the effective magnetic mass $m_{B}$.
Hence, the ratio of the binding energies of the magnetobiexciton and magnetoexciton is
$E_{bex}/E_{ex} = 0.04 \ll 1$  at
$D \gg a_{ex}$, and the
 ratio of radii of the magnetoexciton and magnetobiexciton is
 $r_{0}/a_{ex} = 0.67 (8r^{ex})^{1/4}D^{-1/4} \ll 1$.
So the adiabatic condition is valid.

The magnetobiexciton considered above has only non-zero quadrupole moment  $Q = 3eD^{2}$ (the
large axis of the quadrupole is normal to QWs/or GLs). So
indirect magnetobiexcitons interact at long distances $R \gg D$ as parallel
quadrupoles: $U(R) = 9e^{2}D^{4}/(\epsilon R^{5})$.

\begin{figure}[tbp]
\includegraphics[width = 2.2in]{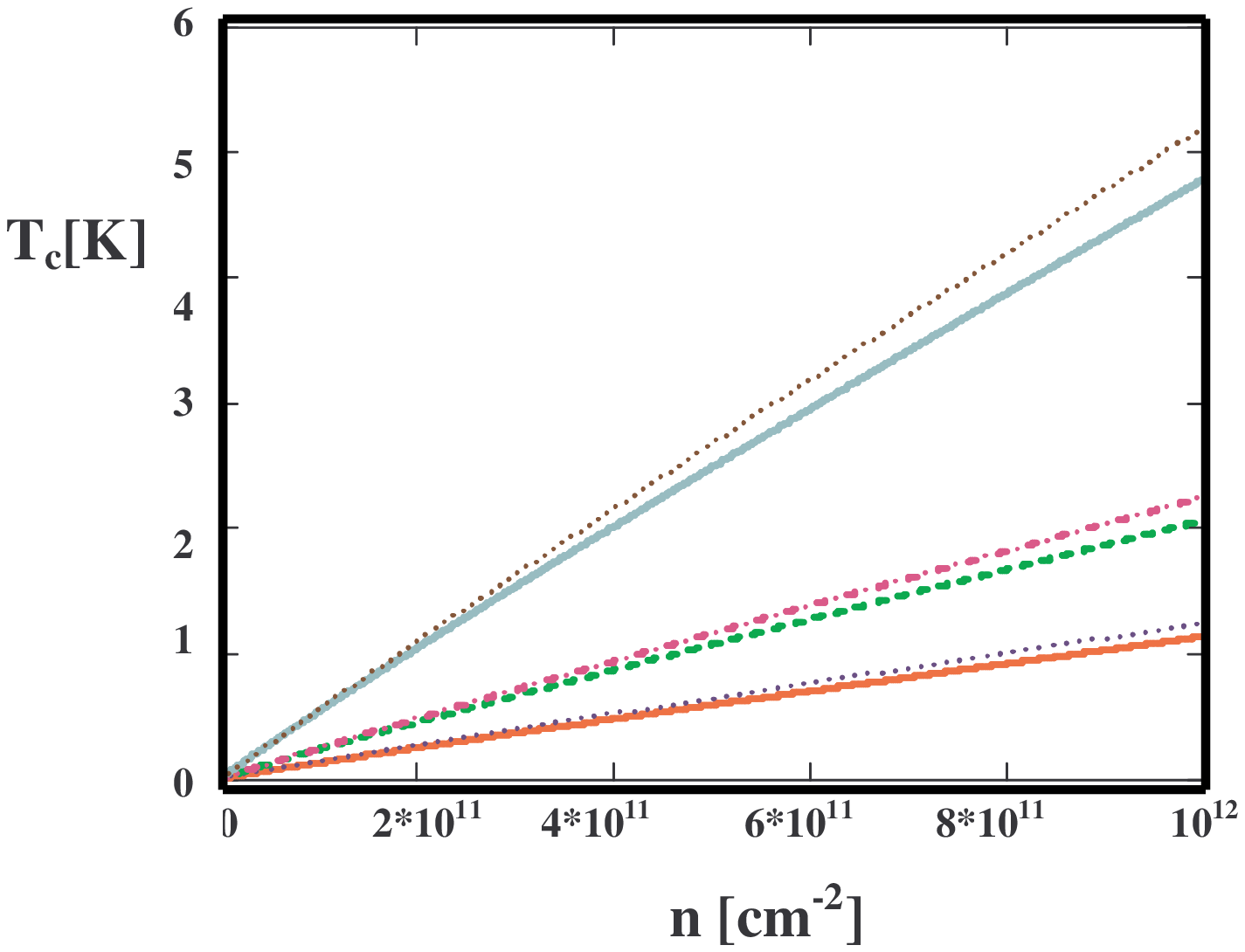}
\caption{Dependence of temperature of the Kosterlitz-Thouless transition $T_c =
T_c (B)$ for the superlattice consisting of QWs
for $GaAs/AlGaAs$: $\protect\epsilon = 13$; and for GLs
separated by the layer of $SiO_{2}$ with $\protect\epsilon = 4.5$ on the
magnetoexciton density $n$ at
$D = 10 nm$ at different magnetic fields. The solid, dashed and thin solid curves for QWs, dotted, dashed-dotted and
 thin dotted curves for GL at $B$: $B = 20 T$, $B = 15 T$ and $B = 10 T$ respectively.}
\label{THD}
\end{figure}

In a 2D system, superfluidity of magnetobiexcitons appears below the
Kosterlitz-Thouless transition temperature $T_{c}=\pi n_{S}(T)/(2m_{B}^{b})$
    ($n_{S}(T)$ is the density of the superfluid component), where only coupled vortices are present \cite{Kosterlitz}. The dependence
of $T_{c}$ on the density of magnetoexcitons at different magnetic field $B$
for superlattice consisting of QWs and GLs was calculated in the ladder approximation for the dilute magnetobiexciton gas, and the results are
represented on Fig.~\ref{THD}.

Thus, it is shown that the low-density system of indirect
magnetoexcitons in a slab of superlattice of second type or consisting of
QWs or GLs in high magnetic field occur to be \textit{%
instable} due to the attraction of magnetoexcitons with opposite dipoles at
large distances. Note that in spite of both QW and graphene realizations represented by completely different Hamiltonians, the effective Hamiltonian in a strong magnetic field was obtained to be the same.  Moreover, for $N$ excitons we have reduced the number of the degrees of freedom from $2N\times 2$ to $N\times 2$  by integrating over the coordinates of the relative motion of e and h. The instability of the ground state of the system of interacting
two-dimensional indirect magnetoexcitons in a slab of superlattice with
alternating electron and hole layers of both QWs and GLs in high magnetic field is claimed due to the attraction between the indirect excitons with opposite directed dipole moments. The stable
system of  indirect quasi-two-dimensional magnetobiexcitons, consisting
from indirect excitons with opposite directed dipole moments is
considered. The low-density system of indirect magnetobiexcitons in
superlattices is \textit{stable} due to the quadrupole-quadrupole repulsion.
So at the pumping increase at low temperatures the excitonic line must
vanish and only magnetobiexcitonic line survives.


\begin{thebibliography}{99}

\bibitem{Snoke} D. W. Snoke, Science {\bf 298}, 1368 (2002).

\bibitem{Butov} L. V. Butov, J. Phys.: Condens. Matter {\bf 16},
R1577 (2004).

\bibitem{Eisenstein} J. P. Eisenstein and A. H. MacDonald,
Nature {\bf 432}, 691 (2004).

\bibitem{Lozovik} Yu. E. Lozovik and V. I. Yudson,
   JETP {\bf 44}, 389
(1976); Physica  {\bf A 93}, 493 (1978).

\bibitem{Birman} J. Zang, D. Schmeltzer and J. L. Birman,
\prl {\bf 71}, 773 (1993).


\bibitem{Littlewood} X. Zhu, P. Littlewood, M. Hybertsen
and T. Rice, \prl {\bf 74}, 1633 (1995).


\bibitem{Vignale}  G. Vignale and A. H. MacDonald,
\prl {\bf 76} 2786 (1996).

\bibitem{Berman}  Yu. E. Lozovik and O. L. Berman,  JETP Lett.
{\bf 64}, 573 (1996);  JETP {\bf 84}, 1027 (1997).



\bibitem{Lerner} I. V. Lerner and Yu. E. Lozovik, JETP {\bf 51}, 588
(1980); JETP, {\bf 53}, 763 (1981).

\bibitem{Paquet} D. Paquet, T. M. Rice, and K. Ueda, \prb
{\bf  32}, 5208 (1985).


\bibitem{Kallin}  C. Kallin and B. I. Halperin,
\prb {\bf 30}, 5655 (1984); \prb {\bf 31}, 3635 (1985).


\bibitem{Yoshioka} D. Yoshioka and A. H. MacDonald,
J. Phys. Soc. Jpn {\bf 59}, 4211 (1990).


\bibitem{Ruvinsky} Yu. E. Lozovik and  A. M. Ruvinsky,
Phys. Lett. {\bf A 227}, 271 (1997); JETP {\bf 85}, 979 (1997).


\bibitem{Ulloa} M. A. Olivares-Robles and S. E. Ulloa, \prb
{\bf 64}, 115302 (2001).


\bibitem{Moskalenko} S. A. Moskalenko, M. A. Liberman, D. W. Snoke, V. V. Botan,  \prb {\bf 66}, 245316 (2002).

\bibitem{Filin} A.~V. Larionov {\it et al.}, JETP Lett. {\bf 75} 570 (2002).

\bibitem{Novoselov2} K. S. Novoselov {\it et al.}, Nature
(London) {\bf 438}, 197 (2005).

\bibitem{Zhang2} Y. B. Zhang {\it et al.}, Nature (London)
{\bf 438}, 201 (2005).

\bibitem{Nomura} K. Nomura and A. H. MacDonald, \prl {\bf 96}, 256602 (2006).

\bibitem{Jain} C. T\H{o}ke, P. E. Lammert, V. H. Crespi, and J. K. Jain,
\prb {\bf 74}, 235417 (2006).

\bibitem{DasSarma} S. Das Sarma, E. H. Hwang, and W.- K. Tse, \prb {\bf 75},
 121406(R) (2007).


\bibitem{Schedin} F. Schedin {\it et al.}, Nature Materials {\bf 6}, 652 (2007).


\bibitem{Kosterlitz} J.~M. Kosterlitz and D.~J. Thouless, J.~ Phys. {\bf C 6},
 1181 (1973); D.~R. Nelson and J.~M. Kosterlitz,
  \prl {\bf 39}, 1201 (1977).

\bibitem{Iyengar} A. Iyengar, J. Wang, H. A. Fertig, and L. Brey,
\prb {\bf 75}, 125430 (2007).

  \bibitem{Gorkov} L. P. Gorkov and I. E. Dzyaloshinskii, JETP {\bf 26}, 449
(1967).

\bibitem{Lukose} V. Lukose, R. Shankar, and G. Baskaran, \prl {\bf 98}, 116802 (2007).

\bibitem{Berman_Lozovik_Gumbs} O. L. Berman, Yu. E. Lozovik, and G. Gumbs,  (cond-mat/0706.0244).



\end{thebibliography}
\end{document}